\documentclass[a4paper,12pt]{article}
\usepackage{latexsym}
\usepackage{amsmath}
\usepackage{amsfonts}
\usepackage{amssymb}
\usepackage{graphicx}

\begin{document}
\title{Human activity modeling and Barabasi's queueing systems}

\author{Ph. Blanchard \\ Fak. f{\"u}r Physik and BiBoS - Univ. Bielefeld \\
D-33615 Bielefeld\\and \\ M.-O. Hongler\\STI/IPR/EPFL\\CH-1015
Lausanne}

\date{}
\maketitle

\abstract{

\noindent It has  been shown by  A.-L. Barabasi that the priority
based scheduling rules in single stage queuing systems (QS)
generates fat tail behavior for the tasks waiting time
distributions (WTD). Such fat tails are due to  the waiting times
of very low priority tasks which stay unserved almost forever as
the task priority indices (PI) are "frozen in time" (i.e. a task
priority is assigned once for all to each incoming task). Relaxing
the "frozen in time" assumption, this paper studies the new
dynamic behavior expected when the priority of each incoming tasks
is time-dependent (i.e. "aging mechanisms" are allowed). For two
class of models, namely 1) a population  type model with an age
structure and 2) a QS with deadlines assigned to the incoming
tasks which is operated under the "earliest-deadline-first"
policy, we are able to analytically extract some relevant
characteristics of the the tasks waiting time distribution. As the
aging mechanism ultimately assign high priority to any long
waiting tasks, fat tails in the WTD cannot find their origin in
the scheduling rule alone thus showing a fundamental difference
between the present and the A.-L. Barab{\`a}si's class of models.}

\noindent \footnotesize{{\bf Keywords} : queueing systems -
waiting time distributions - fat tails distributions - priority
indices dynamics - "earliest-deadline first" scheduling -
 tasks with deadlines - age structured population models.

 \noindent PACS: 89.75.Da $\quad$
0.2.50.-r}

%\vspace{1cm} \centerline{PACS: 89.75.Da $\quad$ 0.2.50.-r}

%%%%%%%%%%%%%%%%%%%%%%%%%%%%%%%%%%%%%%%%%
\section{Introduction} In  recent contributions A.-L. Barabasi
[Barabasi 2005], [V{\'a}zquez 2005] and [V{\'a}zquez et al.  2006] propose
a simplified model of the human activity dynamics. These authors
view the human activity as a decision based queueing system (QS)
where tasks to be executed arrive (randomly) and accumulate before
a server ${\cal S}$ - here ${\cal S}$ stands for the processing
action of the human operator. The time required to process a task
(i.e. {\it the service time}) is generally drawn from a
probability distribution. In addition to the usual features
inherent to any QS, each incoming task is endowed with a priority
index (PI) which indicates the  urgency to process the job. In
this conceptual setting, [Barabasi 2005], [V{\'a}zquez 2005] and
[V{\'a}zquez et al.  2006] study the dynamics arising when the service
policy is not restricted to the usual first-come-first-serve
(FCFS) rule but follows scheduling policies based on PI's. Under
such priority-based scheduling rules, it is shown that the timing
of the tasks follows fat tails probability distribution, (i.e the
activity of the server exhibits bursts separated by long idle
periods). This "burst" character has to be contrasted with the
ubiquitous Poisson behavior  which arises when tasks are executed
according to FCFS or to  purely random order scheduling rules. In
this general context, we shall distinguish between two types of
dynamics, namely:

\begin{itemize}
    \item[] i) {\it Service policies based on fixed (i.e. frozen in time) priority indices}. This case which
    is considered in [Barabasi 2005], [V{\'a}zquez 2005] and [V{\'a}zquez et al.  2006]
     assumes that the value $a$ of the PI  is fixed once for all. Accordingly, very low
    priority jobs are likely to never be served. To circumvent this
    difficulty  [Barabasi 2005], [V{\'a}zquez 2005] and [V{\'a}zquez et al.  2006] introduce an ad-hoc probability
    factor $0\leq p\leq 1$ in terms of which the limit
    $p \rightarrow  1$ corresponds  to a deterministic scheduling strictly based
    on the
   PI's while in the other limit $p \rightarrow  0$ the purely random scheduling is in use. In this setting,
    the {\it waiting time distribution} (WTD) of the tasks before
    service is shown to asymptotically exhibit a fat tail
    behavior. The main point of the  Barabasi's contribution  is to show that
    {\it PI-based scheduling rules can alone generate fat tails in the WTD of unprocessed jobs}.

    \item[] ii) {\it Service policies based on  time-dependent priority indices}. Here the priority
    index is {\it time-dependent}. This typically models situations where {\it the urgency to
      process a task increases with time} and $a(t)$ will hence be represented by increasing time functions.
     Clearly, scheduling rules based on such time-dependent PI do offer
     new specific  dynamical features. They are
      directly relevant in several contexts such as:

     \begin{itemize}
     \item[]  1) {\it Flexible manufacturing systems with limited ressource}. Here a single server is
    conceived  to process different types of jobs but only  a single type can be
produced at a given time $t$
        (i.e. this is the limited ressource constraint). Accordingly, the basic
problem is to dynamically schedule the production to optimally
match random demand arrivals for each types of items. The dynamic
scheduling can be optimally achieved by using time-dependent
priority indices ({\it the Gittins' indices}) which specify in
real time, which type of production to engage [Dusonchet et al.
2003]. Problems of this type belong to the wider class referred as
the {\it Multi-Armed Bandit Problems} in operations research.

\item[]  2) {\it Tasks with deadlines}.  This situation,
 can be idealized by a queueing system where each incoming item has a deadline before
 which it definitely must be processed, [Lehozcky 1996], [Lehozcky 1997], [Doytchinov et al. 2001]. In this case,
 to be later  discussed in the present paper, we can explicitly derive
 the lead-time profile of the waiting jobs obtained under  several scheduling
rules, including the (optimal) time-dependent priority rule  known
as the {\it earliest-deadline-first} policy.

\item[]  3) {\it Waiting time-dependent feedback queueing
systems}. In queueing networks, priority indices based on the
waiting times can be used to schedule the routing through the
network. For networks with loops, such scheduling policies are
able to generate generically stable oscillations of the
populations contained in the waiting room of the queues, [Filliger
et al. 2005].
     \end{itemize}
\end{itemize}

\noindent In the context of QS, the waiting time probability
distribution (WTD), (i.e. the time the tasks spend in the queue
before being processed) is a central quantity to  characterize the
dynamics. It
  strongly depends on the arrival and service stochastic
processes - in particular to the distributions of the {\it
inter-arrival} and {\it service} time intervals. The first moments
of these distributions, enable to define the  trafic load $\rho :=
{\lambda \over \mu} \geq 0$, (i.e.  the ratio between the mean
service time ${1 \over \mu}$ and the mean arrival time ${1 \over
\lambda}$) and clearly the stability of elementary QS is ensured
when $0 \leq \rho <1$. Focusing on the WTD, [Barabasi 2005],
[V{\'a}zquez 2005] and [V{\'a}zquez et al.  2006] emphasized that heavy
tails in the WTD can  have several origins, three of which are
listed below:

\begin{itemize}
    \item[]  i) the {\it heavy traffic load of the server} which induces large
  "bursty" fluctuations in both the WTD and the busy period (BP) of the
  QS. For QS with feedback control driving the dynamics to heavy
  traffic loads, this allows to generate self-organized critical (SOC) dynamics, [Blanchard et al.
  2004] and the resulting fat tails distribution exhibit a decay following a
  $-3/2$ exponent.

    \item[] ii) the presence of {\it fat tails in the service time distribution} produce fat tails of the
    WTD a property which is here independent of the scheduling
    rule [Boxma et al. 2004]. For completeness, we give her a short review of these recent results in
    Appendix A.

    \item[] iii) priority index scheduling rules as discussed
    in [Barabasi 2005], [V{\'a}zquez 2005] and [V{\'a}zquez et al.  2006].
\end{itemize}

  \noindent
  Our present paper focuses on  the case iii) but contrary to the discussion carried in
  [Barabasi 2005], [V{\'a}zquez 2005] and [V{\'a}zquez et al.  2006],
  we shall here consider the dynamics in presence of {\it age-dependent priority
  indices}. As it could have been expected, these aging mechanisms
  generate new behaviors that will be explicitly discussed for two
  classes of models.

\section{Scheduling based on time dependent priority indices}

\noindent  The most naive approach to discuss the dynamics of QS
with scheduling based on {\it time-dependent priority indices} is
to think of a population model in which the members suffer aging
mechanisms which ultimately  will kill them. Naively, we may
consider the population of a city in which members are either born
in the city or  immigrate into it at a certain age and finally die
in the city. Assuming that the death probability {\it depends on
each individual age}, the study of the age structure of the
population exhibits some of the salient features of our original
QS. This is the class of models to be discussed in section
\ref{POPU}. Later in section \ref{TIMERQ}, we shall return to the
original model of L. Barab{\'a}si and consider a simple QS where each
task waiting to be processed carries a deadline (playing the role
of a PI) and as time flows the these deadlines steadily reduce -
this implies a ({\it time dependence of the PI}). At a given time,
the scheduling of the tasks follows the "earliest-deadline-first"
(EDF) policy and given a queue length configuration, we shall
discuss the lead-time (lead-time = deadline - current time)
profile of the tasks waiting to be served.

\subsection{Tasks population dynamics with time dependent priority indices}\label{POPU}

\noindent Consider a population of tasks waiting to be processed
by ${\cal S}$ with the following characteristics:

\begin{itemize}

\item[] i) an inflow of  new tasks steadily enters into the
queueing system. Each tasks is endowed with a priority index (PI)
$a$ which indicates its degree of urgency to be processed. In
general, the tasks are heterogenous as the PI are different. In
the time interval $[t, t+\Delta t]$, the number of incoming jobs
exhibiting an initial PI in the interval $[a, a + \Delta a]$ is
characterized by $G(a,t) \Delta t \Delta a$.

\item[] ii) Contrary to the
    situations  discussed in [Barabasi 2005], an "aging" process
    directly affects the urgency to process a given task. In other
    words the priority index $a$ is not frozen in time but $a=a(t)$
    monotonously increases with time $t$. For an infinitesimal time increase $\Delta t$, in the simplest
    case  we shall have
    $a(t+ \Delta t) = a(t) +  \Delta t$. Here we slightly  generalize this and allow inhomogeneous aging
    rates written as $p(a)>0$ meaning that $a( t+ \Delta t) = a(t)
    + p(a) \Delta t$.

\item[] iii) the scheduling policy  depends on the PI of the tasks
in the queue and we will focus on the natural policy {\it "process
the highest PI first"}.

    \item[] iv) at time $t$, a scalar field $M(a,t)$ counts
 the number of waiting tasks with priority
index $a$. Hence $M(a,t) \Delta a$ is the number with PI $p(a) \in
[a + \Delta a]$. Accordingly, the total workload facing the human
server $\cal {S}$ will be given, at time $t$ by:

\begin{equation}\label{WL}
    L(t) = \int_{0}^{\infty} M(a,t) \,da.
\end{equation}

\item[]v)  in the time interval  $[t, t+\Delta t]$, the server
$\cal{S}$  processes tasks with an $a$-dependent rate $\mu(a)
\Delta t$. Typically $\mu(a)$ could be a monotonously increasing
function of $a$. As the service rate $\mu(a)$ explicitly depend on
the PI $a$, it therefore plays an effective role of service
discipline.

\end{itemize}

\noindent
 The previous elementary hypotheses imply an evolution in the form:

  $$M(a + p(a)\Delta t, t + \Delta t) \Delta a \approx
    M(a,t) \Delta a- \mu(a) M(a,t) \Delta a \Delta t + G(a,t)\Delta t \Delta a$$

    \noindent
Dividing by $\Delta a  \Delta t$, we end, in the limits $\Delta a
\rightarrow 0$ and $\Delta t \rightarrow 0$,  with the scalar
linear field  equation:

\begin{equation}\label{EVOL}
    {\partial \over \partial t}M(a,t) + p(a){\partial \over \partial
    a}M(a,t)+ \mu(a) M(a,t) = G(a,t).
\end{equation}

\noindent It is worth to remark that the dynamics given by
Eq.(\ref{EVOL}) is closely related to the  famous McKendrick's age
structured population dynamics, [Brauer et al. 2001].

\noindent Assuming stationarity for the incoming flow of tasks
(i.e. $G(a,t)= G_{s}(a)$), the linearity of Eq.(\ref{EVOL})
enables to explicitly write its stationary solution as:

\begin{equation}\label{STA}
    M(a) = \pi(a) \left[C + \int_{0}^{a}  {G_{s}(z) \over p(z) \,
    \pi(z)}\, dz\right],
\end{equation}

\noindent where
\begin{equation}\label{pi}
    \pi(z) = \exp\left\{-\int^{z}{\mu(y) \over
    p(y)}\, dy\right\},
\end{equation}

\noindent with an integration constant $C < \infty$ remaining yet
to be determined. Assume that the  PI attached to the incoming
jobs do not exceed a limiting value $T$, namely:
\begin{equation}\label{T}
    G(a,t) = \mathbb{I}(a < T) \hat{G}(a,t)\,\,\,\,\,\,\,\,\,\, \Rightarrow \,\,\,\,\,\,\,\,\,\,
    G_{s}(a) = \mathbb{I}(a < T) \hat{G}_{s}(a),
\end{equation}

\noindent where $\mathbb{I}(a < T)$ is the indicator function. In
other words Eq.(\ref{T})  indicates that the new coming jobs do
not exhibit arbitrarily high PI's.

 \noindent This enables to define:

\begin{equation}\label{NOTA}
\Psi(T) := \int_{0}^{\infty} {G_{s}(z) \over p(z) \,
    \pi(z)}\, dz = \int_{0}^{T} {\hat{G}_{s}(z) \over p(z) \,
    \pi(z)}\, dz
\end{equation}

\noindent and Eq.(\ref{STA}) reads as:

\begin{equation}\label{FINASOL}
M(a) = \left\{\begin{array}{c}  \pi(a) \left[C + \int_{0}^{a}
{\hat{G}_{s}(z) \over p(z) \,
    \pi(z)}\, dz\right]\qquad\mbox{if} \qquad a\leq T
    \\ \pi(a) \left[C+ \psi(T)\right]\qquad \qquad \qquad\mbox{if} \qquad a>
    T.
\end{array}\right.
\end{equation}

\noindent The asymptotic behavior  of $M(a)$ for  $a \rightarrow
\infty$ is entirely due to $\pi(a)$, (the square bracket terms are
bounded by constants) and therefore Eqs. (\ref{pi}) and
(\ref{FINASOL}) imply:

\begin{equation}\label{ASY}
    M(a) \approx \pi(a) \approx
    \left\{\begin{array}{c}
e^{{-k \over q} a^{q}}\qquad\mbox{when} \qquad {\mu(a)\over p(a)}
\varpropto ka^{q-1}\qquad\mbox{with} \qquad q >  0,\\
{1\over a^{k}}\qquad\,\,\,\,\,\,\,\,\,\mbox{when} \qquad
{\mu(a)\over p(a)} \varpropto {k \over a}, \qquad \qquad
\qquad\qquad\qquad \\
\end{array}
    \right.
\end{equation}

\noindent In view of Eq.(\ref{ASY}), the following alternatives
occur:

\begin{itemize}
    \item[] a) for  $q < 0$ in Eq.(\ref{ASY}), the
integral $\int_{0}^{\infty}M(z)\, dz$ does not exist. In this case
an ever growing population of tasks accumulates in front of the
server and the queueing process is exploding.

    \item[] b) for $q > 0$, a stationary regime exists and in this case the constant
$C < \infty$ in Eq.(\ref{FINASOL}) can be determined by solving:

\begin{equation}\label{BALANCE}
    \int_{0}^{\infty} G_{s}(z) \, dz = \int_{0}^{\infty} M(z)
    \mu(z) \,dz,
\end{equation}

\noindent which  expresses a global balance between the stationary
incoming and out going flows of tasks.

    \item[] c) for $q=0$ which implies that ${\mu(a)\over p(a)} \varpropto {k \over a}$,
Eq.(\ref{ASY}) {\it produces an exponent-$k$ fat tail distribution
for $M(a)$ counting the number of waiting tasks with PI $a$} in
the system. For $T < \infty$ and $ a \rightarrow \infty$, the fat
tail of $M(a)$  is populated by long waiting  tasks i.e. those
having spent more than $a-T$ waiting inside the system before
being served. In the limiting case, for which $\mu(a) = \mu = {\rm
const}$ and $p(a) =a$ (i.e. aging directly proportional to time)
which leads to $q=0$ in Eq.(\ref{ASY}), the density $M(a)$
coincides directly with the WTD for $a \rightarrow \infty$.
\end{itemize}

\noindent This population model shares  several features  with the
Barab{\'a}si's model, namely:

\begin{itemize}
    \item[] a) when a stationary regime exists, the function
$\hat{G}_{s}(a)$ which  here plays the role of the initial PI
distribution in [Barabasi 2005], [V{\'a}zquez 2005] and [V{\'a}zquez et
al. 2006], does not affect the tail behavior given by
Eq.(\ref{ASY}).

    \item[] b)  the scheduling rule here is implicitly governed  by the service rate
    $\mu(a)$ which itself depend on time as the PI $a=a(t)$ are time-dependent.
    Note that $\mu(a)$ directly influences the asymptotic behavior of Eq.(\ref{ASY}). In particular for case
    c), the tail exponent explicitly depends on  $\mu(a)$.

\end{itemize}

\noindent Besides  the similarities, we now also point out the
important differences between the present population model and the
model discussed in  [Barabasi 2005], [V{\'a}zquez 2005] and [V{\'a}zquez
et al. 2006]:

\begin{itemize}

    \item[] a) the service is not restricted to a single task at
a given time (i.e. the service ressource is not limited). Indeed
$\mu (a)$ describes an average flow of service and hence several
tasks can be processed simultaneously - (in the city population
model the service corresponds to death and several individual may
die simultaneously).

    \item[] b) while the fat tail in [Barabasi 2005], [V{\'a}zquez 2005] and
[V{\'a}zquez et al. 2006] is entirely due to the scheduling rule and
therefore occurs even for QS far from traffic saturation, this is
not so in the population model. Indeed in this last case, fat
tails are due to heavy traffic loads occurring when the flow of
incoming tasks nearly saturates the server, (this is implied by
$q=0$ in Eq.(\ref{ASY})) - for lower
 loads occurring when $q>0$  the fat tail in Eq.(\ref{ASY})
 disappears.

\end{itemize}

%%%%%%%%%%%%%%%%%%%%%%%%%%%%%%%%%%%%%%%%%%%%%%%%%%%%%%%%%%%%%%%%%%%%%%%%%%%%%%%%%%%%%%%%%%%%%%%%%%%%%%%%%%%%%
%%%%%%%%%%%%%%%%%%%%%%%%%%%%%%%%%%%%%%%%%%%%%%%%%%%%%%%%%%%%%%%%%%%%%%%%%%%%%%%%%%%%%%%%%%%%%%%%%%%%%%%%%%%%%
%%%%%%%%%%%%%%%%%%%%%%%%%%%%%%%%%%%%%%%%%%%%%%%%%%%%%%%%%%%%%%%%%%%%%%%%%%%%%%%%%%%%%%%%%%%%%%%%%%%%%%%%%%%%%
%%%%%%%%%%%%%%%%%%%%%%%%%%%%%%%%%%%%%%%%%%%%%%%%%%%%%%%%%%%%%%%%%%%%%%%%%%%%%%%%%%%%%%%%%%%%%%%%%%%%%%%%%%%%%
%%%%%%%%%%%%%%%%%%%%%%%%%%%%%%%%%%%%%%%%%%%%%%%%%%%%%%%%%%%%%%%%%%%%%%%%%%%%%%%%%%%%%%%%%%%%%%%%%%%%%%%%%%%%%
%%%%%%%%%%%%%%%%%%%%%%%%%%%%%%%%%%%%%%%%%%%%%%%%%%%%%%%%%%%%%%%%%%%%%%%%%%%%%%%%%%%%%%%%%%%%%%%%%%%%%%%%%%%%%
%%%%%%%%%%%%%%%%%%%%%%%%%%%%%%%%%%%%%%%%%%%%%%%%%%%%%%%%%%%%%%%%%%%%%%%%%%%%%%%%%%%%%%%%%%%%%%%%%%%%%%%%%%%%%
%%%%%%%%%%%%%%%%%%%%%%%%%%%%%%%%%%%%%%%%%%%%%%%%%%%%%%%%%%%%%%%%%%%%%%%%%%%%%%%%%%%%%%%%%%%%%%%%%%%%%%%%%%%%%
%%%%%%%%%%%%%%%%%%%%%%%%%%%%%%%%%%%%%%%%%%%%%%%%%%%%%%%%%%%%%%%%%%%%%%%%%%%%%%%%%%%%%%%%%%%%%%%%%%%%%%%%%%%%%
%%%%%%%%%%%%%%%%%%%%%%%%%%%%%%%%%%%%%%%%%%%%%%%%%%%%%%%%%%%%%%%%%%%%%%%%%%%%%%%%%%%%%%%%%%%%%%%%%%%%%%%%%%%%%
%%%%%%%%%%%%%%%%%%%%%%%%%%%%%%%%%%%%%%%%%%%%%%%%%%%%%%%%%%%%%%%%%%%%%%%%%%%%%%%%%%%%%%%%%%%%%%%%%%%%%%%%%%%%%
%%%%%%%%%%%%%%%%%%%%%%%%%%%%%%%%%%%%%%%%%%%%%%%%%%%%%%%%%%%%%%%%%%%%%%%%%%%%%%%%%%%%%%%%%%%%%%%%%%%%%%%%%%%%%
%%%%%%%%%%%%%%%%%%%%%%%%%%%%%%%%%%%%%%%%%%%%%%%%%%%%%%%%%%%%%%%%%%%%%%%%%%%%%%%%%%%%%%%%%%%%%%%%%%%%%%%%%%%%%

\vspace{1cm} \noindent

\subsection{Stochastic dynamics. Real-time queueing dynamics}\label{TIMERQ}

\noindent In this section we will use  the results of the
real-time queueing theory (RTQS), pioneered in [Lehozcky 1996], to
explore situations where the incoming jobs have a deadline - this
problem is already suggested in [Barabasi 2005]. Based on
[Lehozcky 1996], [Lehozcky 1997], [Baldwin et al. 2000] and
[Doytchinov et al. 2001],  first recall the basic hypotheses and
the relevant results of RTQS's. Consider a general single server
QS with arrival and service being described by independent renewal
processes with average ${1 \over \lambda}$ respectively ${1 \over
\mu}$ and {\bf finite variances} for both renewal processes. Each
incoming task arrives with a random hard time relative deadline
${\cal D}$ drawn from a PDF $G(x)$ with a density $g(x)$:

$$
{\rm Prob}\left\{ 0\leq {\cal D} \leq x \right\} = G(x),
$$

\noindent with average $\langle {\cal D } \rangle$:

$$
\langle{\cal D}\rangle = \int_{0}^{\infty}\left(1 -
    G(x)\right)dx = \int_{0}^{\infty}x g(x)dx.
$$

\noindent At a given time $t$, we define the lead time ${\cal L}$
to be given by:

\begin{equation}\label{LTIME}
    {\cal L} = {\cal D} - t,
\end{equation}

\noindent Assume now that the lead time ${\cal L}$ plays the role
of a priority index and  the service is delivered by using the
{\it earliest-deadline-first} (EDF) rule with preemption (i.e.
the server always processes the job with the shortest lead time
${\cal L}$). Preemption implies that whenever an incoming job
exhibits  a shorter ${\cal L}$ than the one currently in service,
this incoming job is processed before, (i.e. {\it preempts}), the
 currently engaged task which service is postponed. The EDF rule
 directly
 corresponds  to the deterministic policy (i.e. $p=0, \gamma= 0$
 in the original Barab{\'a}si's contribution [Barab{\'a}si 2005].

 \noindent At a given
time, one can define a probability distribution corresponding to
the {\it lead time profile} (LTP), ${\rm Prob}\left\{-\infty\leq
{\cal L} \leq x\right\}:=F(x)$, of the jobs waiting in the QS. The
LTP specifies the repartition of tasks having a given ${\cal L}$
at time $t$. Knowing the queueing population $Q$ at a given time,
it is shown in [Doytchinov et al. 2001] that for heavy traffic
regimes, the LTP can, in a first order approximation scheme,
expressed by a simple analytical form. Specifically, following
[Doytchinov et al. 2001], define a frontier ${\cal F}(Q)>0$ to be
the unique solution of the equation:

\begin{equation}\label{FRONTIER}
    {Q \over \lambda} =\int_{{\cal F}(Q)}^{\infty}\left(1 -
    G(x)\right)dx, \qquad \quad
    (x \in \mathbb{R} \,\,\, {\rm and } \,\,\, G(x) \equiv 0 \,\,\, {\rm for} \,\,\,  x<0).
\end{equation}

\begin{figure}[htbp]
     \centerline{\includegraphics[height=3.5cm]{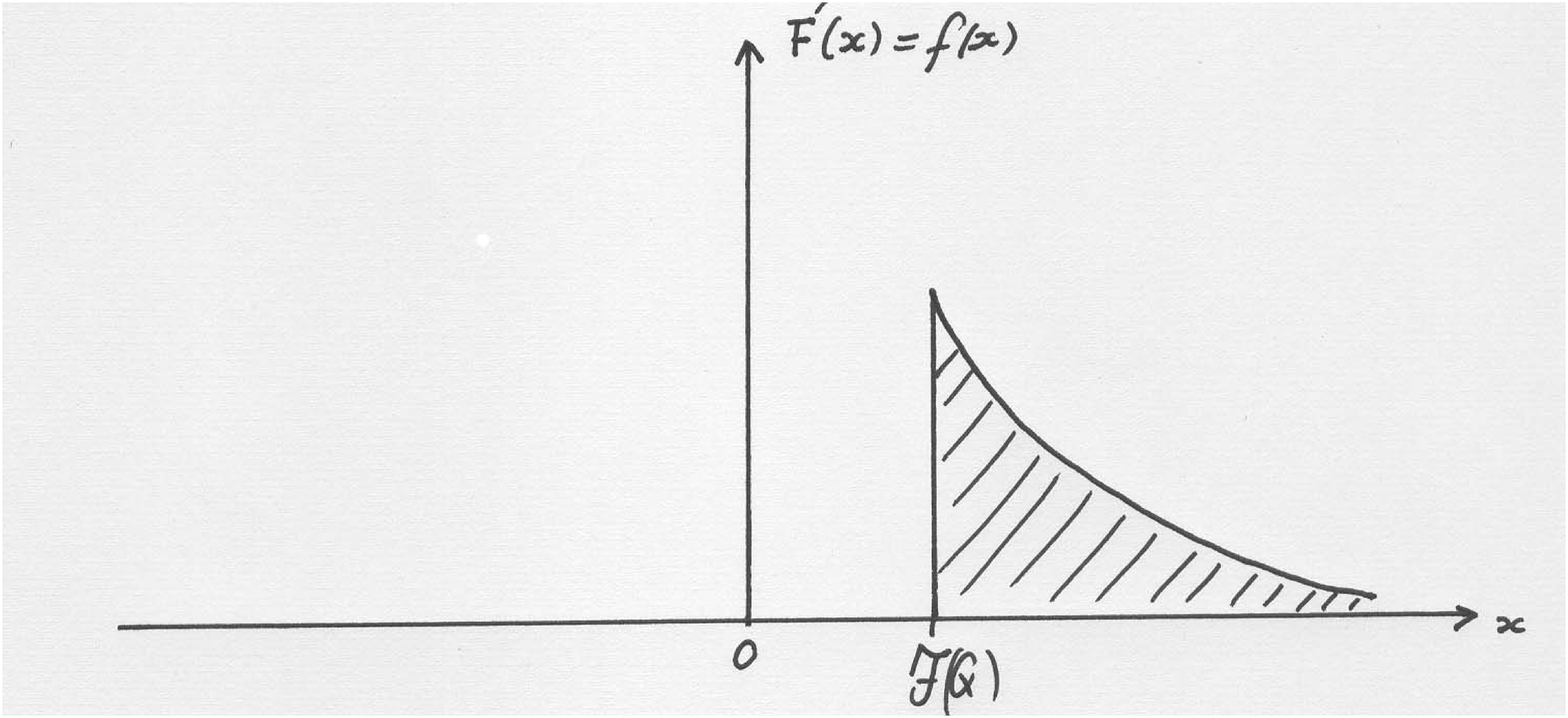}}
     \caption{Density of the lead time profile $f(x)  = {dF(x) \over dx}$ when ${\cal F}(Q) >0$.}
     \protect\label{FDEQPOS}
 \end{figure}

\begin{figure}[htbp]
     \centerline{\includegraphics[height=4cm]{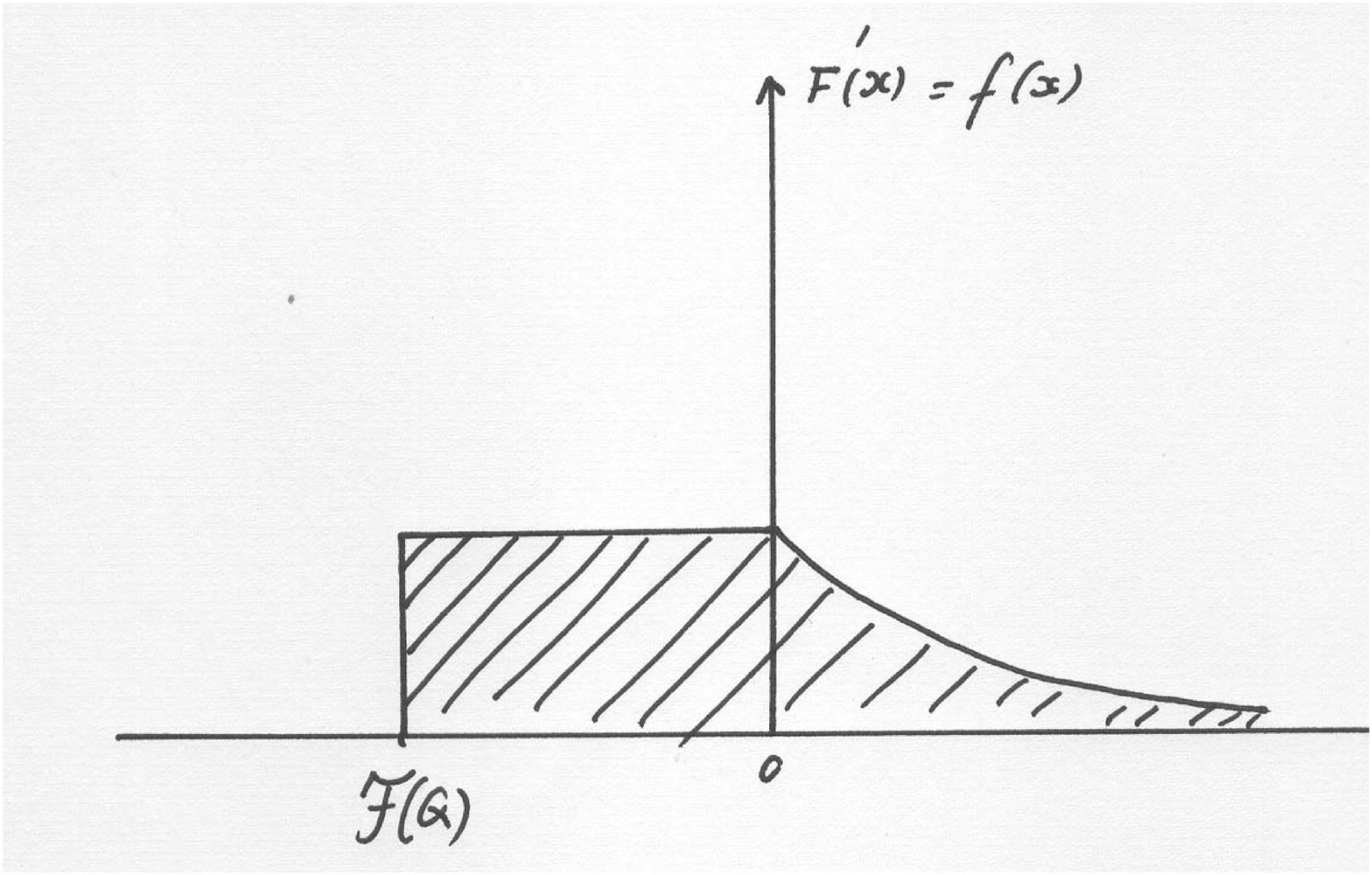}}
     \caption{Density of the lead time profile $f(x) = {dF(x) \over dx}$ when ${\cal F}(Q) <0.$}
     \protect\label{FDEQNEG}
 \end{figure}

\noindent In [Doytchinov et al.
 2001], it is shown that two alternative regimes can occur:

\begin{itemize}
    \item[] a) {\it Jobs served before deadline}. Solving Eq.(\ref{FRONTIER}), a positive value ${\cal F}(Q) >0$ is obtained. In this case, the LTP
    cumulative distribution $F(x)$ takes the form,  see Figure \ref{FDEQPOS}:

\begin{equation}\label{LTP1}
F(x) = \left\{
\begin{array}{l@{\qquad}l} 0 & {\rm when}\ x < {\cal F}(Q),\\ \, \\
{\lambda \over Q}\left(\int_{x}^{\infty}\left[1-G(\eta)\right]
d\eta \right)& {\rm when}\ {\cal F}(Q) \leq x.
\\
\end{array}
    \right.
\end{equation}

    \item[] b) {\it Jobs served after  deadline}. Solving Eq.(\ref{FRONTIER}), a negative value  ${\cal F}(Q) <0 $
results. In this case, the LTP
   cumulative distribution $F(x)$ takes the form, see Figure \ref{FDEQNEG}:

\begin{equation}\label{LTP2}
F(x) = \left\{\begin{array} {l@{\qquad}l} 0 &{\rm when}\ x< {\cal
F} (Q) = \left\langle{\cal D} \right\rangle -{Q
\over \lambda}<0,\\ \, \\
   q \left[x - \langle{\cal D}\rangle + {Q \over\lambda } \right]
& {\rm when} \
 {\cal F}(Q) \leq x <0,\\ \, \\
 {\lambda \over Q}\left\{\int_{x}^{\infty}\left[1-G(\eta)\right]d\eta\right\}& {\rm when} \ 0 \leq
 x.\\
\end{array}
    \right.
\end{equation}

\noindent where $q = {{\lambda \over Q} \langle{\cal D}\rangle
\over{Q\over \lambda} -\langle{\cal D}\rangle}$.
\end{itemize}

\vspace{1cm}
 \noindent {\bf Remark}. The alternative regimes given by Eqs.(\ref{LTP1}) and (\ref{LTP2})
  can be heuristically understood by invoking the  {\it Little law} which connects the average
queue length $\langle Q \rangle$ with the average waiting time
$\langle W \rangle$,  [Cohen 73]:

\begin{equation}\label{LITTLE}
\langle Q \rangle = \lambda \, \langle W \rangle,
\end{equation}

\noindent which is independent of the scheduling policy. In view
of Eqs.(\ref{FRONTIER}) and (\ref{LITTLE}), one obviously suspects
that the $LTP$  strongly
 depends on the sign of the difference $\langle{\cal
 D}\rangle-{\langle Q \rangle \over \lambda}= \langle{\cal D}\rangle- \langle W\rangle$.
Intuitively, when $\langle W\rangle$ exceeds $\langle D \rangle$,
it is
 expected, in the average, that processed jobs will be delivered too late and conversely.
 While the above heuristic arguments is strictly valid only for the averages, [Doytchinov et al.
 2001],  were able to show that in heavy traffic regimes, it also holds also for the LTP
 given in Eqs.(\ref{LTP1}) and (\ref{LTP2}).

\vspace{1cm}  \noindent Assuming that the arriving tasks have
positive deadlines, the LTP as given by Eqs.(\ref{LTP1}) and
(\ref{LTP2}) imply:

\begin{itemize}
    \item[] a) If the left-hand support of the LTP is negative,
    then a job entering into service is already late, (case of Eq.(\ref{LTP2})) see Figure \ref{FDEQNEG}.
    \item[] b) If the left-hand support of the LTP is positive
    then a   jobs enters into service with a positive lead time, (case of Eq.(\ref{LTP1}))  see Figure
    \ref{FDEQPOS}. Accordingly, it is likely that the tasks will
    be completed before the deadline expired.

    \item[] c) The critical value $Q^{*}={\left\langle{\cal D} \right\rangle\over \lambda}$  for which ${\cal
    F}(Q)=0$, corresponds to a queue length for which customers
    are likely to become  late. Choosing $Q$ exactly to $Q^{*}$, we cannot expect lateness to
    disappear completely but for $ Q < Q^{*}$ lateness will be strongly
    reduced a behavior clearly confirmed by  simulation
    experiments [Doytchinov et al.
 2001] and [Baldwin et al. 2000].

 \item[] d)  For deadline distributions $G(x)$ with fat tails, it
 follows immediately from Eqs.(\ref{LTP1}) and (\ref{LTP2}) that
 the LTP does possess a fat tail.

\end{itemize}

\subsubsection{"First come  first served" (FCFS)  scheduling  policies}
    \noindent
 For the choice $g(x) = \delta(x)$,
(i.e. zero deadline), the EDF scheduling policy directly coincides
 with  the FCFS rule. Indeed, in this case ${\cal F}(Q)=0$ and the LTP
 density is given by  Eq.(\ref{LTP1})  is  a uniform probability density
$U\left[- {Q\over \lambda},0\right]$,
 ($\left[- {Q\over \lambda},0\right]$ being its support). This
 expresses the fact that in the heavy traffic regime $\rho = \lambda / \mu \approx 1$, the waiting
 time behaves as $Q \times ({1\over \mu}) \approx Q \times ({1\over
 \lambda})$  leading to a LPT linearly growing
 with $Q$. For  general $G(x)$, the LTP associated with a FCFS scheduling rule will
  be  given by the convolution of the deadline distribution $G(x)$
 with the uniform distribution $U\left[- {Q\over
 \lambda},0\right]$. Indeed, adding the task deadlines with the
 time spent in t5he queue,  we recover the tasks lead-time. Accordingly, in the heavy traffic regime and for a given queue
length $Q$, one  explicitly knows the LTP's for both the EDF and
the FCFS scheduling policies thus enabling to explicitly
appreciate
 their respective characteristics. In particular, using
Eqs.(\ref{LTP1}) and (\ref{LTP2}), one can conclude that for a
given queue length $Q$ the FCFS scheduling rule  the associated
LTP $F(x)$ being the convolution of $G(x)$ with the $U\left[-
{Q\over
    \lambda},0\right]$, it takes the form:

\begin{equation}\label{CONVOLO}
    F(x) = \left\{
\begin{array}{l@{\qquad}l}
0 & {\rm when}\  x < -{Q\over \lambda},\\
\,
\\  {\lambda \over Q}\int_{-\left({Q \over \lambda}\right)}^{x}\left[G( \xi + {Q \over
\lambda})\right]d\xi &{\rm when} -{Q \over \lambda}\leq x < 0, \\
\,
\\  \kappa  +{\lambda \over Q}
\left\{\int_{0}^{x}\left[G( \xi + {Q \over \lambda}) - G(\xi)\right] d\xi \right\}&{\rm when}\ x\geq 0,\\
\end{array}
    \right.
\end{equation}

\noindent where the constant $\kappa$ reads as:
$$\kappa = {\lambda \over Q}\int_{-\left({Q \over \lambda}\right)}^{0}\left[G(\xi + {Q \over
\lambda})\right]d\xi. $$

\noindent  Eq.(\ref{CONVOLO}) allows to emphasize the following
features:

\begin{itemize}
    \item[] i) When the left-hand support of the deadline distribution $G(x)$ is
    larger than ${Q \over \lambda}$, the left boundary of the support of $F(x)$ is
    larger than $0$ and therefore the jobs experience no delay
    when entering into service.

    \item[] ii) If the left-hand  support of $G(x)$ is
    smaller than ${Q \over \lambda}$, then it may happen that
     the LTP exhibits a negative left-hand support under the FCFS
     policy and a positive left-hand support under the  EDF
    scheduling rule. Hence in this last situation, the FCFS policy would deliver tasks with lateness
    while the EDF  tasks will be processed in due time.
     This explicitly confirms intuition that  EDF is indeed an efficient
    policy. It has been shown that the  EDF
scheduling rule is optimal for minimizing the number of jobs
processed after the deadline [Panwar et al. 88].

    \item[] iii) If $G(x)$ exhibits a fat tail for $x \rightarrow \infty$ so has the
    LTP and this whatever the scheduling rule in use. This can e
    directly verified from Eq.(\ref{CONVOLO}) by studying the LTP
    density $f(x) = {d F(x) \over dx} $ for $x \rightarrow
    \infty$, we have:

    $$
    f(x) = {\lambda \over Q}\left[ G(x+ {Q \over \lambda}) - G(x) \right], \qquad \qquad
    {\rm for } \qquad  x \rightarrow \infty,
$$
\noindent which when  $G(x) \sim 1- x^{-q}$ and for ${Q \over
\lambda} < {\rm const}$ takes the form

\begin{equation}\label{LTP-DENSSS}
f(x) \sim x^{-(q+1)}, \qquad \qquad
    {\rm for } \qquad  x  \rightarrow \infty.
\end{equation}

    Hence, the LTP  inherits the fat tail property of
    $G(x)$ and this even when using the optimal EDF scheduling
    rule - a fully explicit illustration involving the Pareto probability distribution is given
    in the Appendix B.

\end{itemize}

\noindent The results obtained for the LTP, enable to get the
asymptotic properties of the  waiting time distribution WTD.
Indeed, under the EDF policy, the more urgent jobs are served
first and therefore the waiting time before service of the
queueing jobs will be larger or equal to the lead-time.
Accordingly, if the LTP exhibits a fat tail distribution so will
the WTD. Hence, while the EDF policy  decreases, compared with the
FCFS rule, the number of jobs served after their deadline, it
cannot get rid of the fat tails generated by the deadline
probability distribution $G(x)$. Let us emphasize here, the fat
tails of the LTP (and hence of the WTD) are here entirely due to
$G(x)$ and his asymptotic behavior of the LTP is shared by both
the EDF and FCFS policies. This is fundamentally different from
the frozen in time PI models discussed in [Barabasi 2005],
[V{\'a}zquez 2005] and [V{\'a}zquez et al. 2006] where the fat tail
behavior does not depend on $G(x)$. This
 can be heuristically understood as, in [Barabasi 2005], [V{\'a}zquez
2005] and [V{\'a}zquez et al.  2006], the fat tail is mainly due to
the low priority jobs which, as no aging mechanism occurs,  are
likely to never be served. Note that in [Barabasi 2005], [V{\'a}zquez
2005] and [V{\'a}zquez et al.  2006], stable queueing models, (i.e.
those for which the traffic $\rho <1$),   fat tails of the WTD
disappear under a FCFS scheduling rule. Indeed without priority
scheduling, the WTD always follows an exponential asymptotic
decaying behavior. In presence of time dependent PI, all tasks do
finally acquire a high priority and this aging  mechanism
precludes the formation of a fat tail solely due to the scheduling
rule. Accordingly, in presence of aging PI, the generation of WTD
with fat tails will  be due to $G(x)$.

\section{Conclusion and summary}

\noindent There are several possibilities to analytically discuss
the scheduling of tasks in QS with time dependent priority indices
and to infer on the existence of fat tails for  the asymptotic
behavior of the resulting WTD. In this note, we propose two
distinct models where an explicit analysis can be developed. Our
first model is directly inspired by the study of age classes in
population dynamics where the mortality rate increases with the
age of the individuals. In this context, identifying the service
of the QS with the death of an individual, this dynamics is
closely related to the scheduling based on PI, the indices here
being the age of the individuals and the immigration with
different ages plays the role of incoming tasks with different
priorities. For this class of dynamics, it is straightforward to
show that fat tails of the WTD can develop on the onset of
stability of the population model. As in the original Barab{\'a}si
model, the tail behavior of the WTD does not depend on the detail
nature of the PI but only on the scheduling rule - (corresponding
in the population model to the mortality rate). Our second
modeling frame which is closer to the Barab{\'a}si original idea, we
consider a classic QS in which the scheduling rule is based on the
deadlines attached to each incoming tasks. As time flows, the
deadlines reduce and hence the waiting  tasks acquire a higher
priority to be processed. In the heavy traffic limit, i.e. for
regimes where the law of large numbers dominate, it is possible to
analytically derive the lead-time profile (lead-time = deadline
minus the time elapsed in queueing) of the waiting tasks and from
this to get information on the asymptotic behavior of the
associated WTD. In this case and contrary to the conclusions
exposed in [Barabasi 2005], [V{\'a}zquez 2005] and [V{\'a}zquez et al.
2006], the scheduling rule solely cannot generate fat tails in the
WTD. Fat tail in [Barabasi 2005], [V{\'a}zquez 2005] and [V{\'a}zquez et
al. 2006] are due to low priority jobs which are likely to never
be served. This possibility disappears when time-dependent PI are
considered as, due to aging, initially low priority tasks do
acquire, with time, high priorities and hence will not stay
unprocessed forever. This  precludes the formation of fat tails in
the WTD. We finally observe that, in this second class of models,
the only possibility to generate fat tails is to feed the QS with
tasks deadlines drawn from a fat tail distribution.

\vspace{2cm} \noindent {\bf Acknowledgements}. M.O.H. thanks
numerous fruitful discussions with Olivier Gallay and Roger
Filliger.

\newpage

\section*{Appendix  A - Waiting time distributions for QS with fat tail service times}\noindent

\vspace{0.5cm} \noindent Let us reproduce here a result recently
derived in [Boxma et al. 2004]:

\vspace{0.5cm} \noindent \underline{Theorem 1}. Assume that the
(random) service time in a $M/G/1$ QS is drawn from a PDF with a
regularly varying tail at infinity with index $\nu \in (-1, -2)$,
(regularly varying  with index $\nu \in (-1, -2)$ $\Rightarrow$
fat tail with index $\nu \in (-1, -2)$). For this range of
asymptotic behaviors of the PDF, the first moment $\beta$ of the
service exists. Assume further that the service is delivered
according to a random order discipline. Then the waiting time
distribution $W_{{\rm ROS}}$ exhibits a fat tail with index
$\left(1-\nu \right)\in (-1,0)$ and more precisely, we can write:

\begin{equation}\label{ROS}
    {\rm Prob}(W_{{\rm ROS}} > x )\varpropto C{\rho \over 1-\rho}\,
    {h(\rho, \nu) \over \beta \, \Gamma(2-\nu)}\,  x^{1-\nu}\, {\cal
    L}(x),
\end{equation}

\noindent where $\rho <1$ is the traffic intensity, $\beta$ the
average service time, ${\cal L}(x)$ a slowly varying function and

$$
h(\rho, \nu) := \int_{0}^{1} f(u, \rho, \nu) du,
$$

\noindent with:

$$
f(u, \rho, \nu) := { \rho \over 1-\rho}\left({\rho \, u\over
1-\rho}\right)^{\nu-1} \,\left(1-u\right)^{{1 \over 1-\rho}}+
\left(1+ {\rho \, u \over1-\rho} \right)^{\nu}\,
\left(1-u\right)^{{1\over 1-\rho}-1}.
$$

\vspace{1cm} \noindent The fat tail behavior given in
Eq.(\ref{ROS}) is therefore entirely inherited from the  fat tail
behavior of the service and is not affected by any reduction of
the trafic intensity $\rho$. Note also that change of the
scheduling rule cannot get rid of this fat tail behavior. This
point can be explicitly observed in [Cohen 1973], [Pakes 1975] who
show  that for the previous $M/G/1$ QS with a random order service
(ROS) service discipline, one obtains:

\begin{equation}\label{FCFS_ROS}
{\rm Prob}(W_{{\rm ROS}} > x )\varpropto h(\rho,\nu) {\rm
Prob}(W_{{\rm FCFS}} > x ), \,\,\,\,\,\,\,\,\,\,\,{\rm for }
\,\,\, x \rightarrow \infty,
\end{equation}

\noindent from which we directly observe that {\it the fat tail in
the asymptotic behavior in not altered by a change of the
scheduling rule}.

 \noindent Note finally that for the $M/M/1$ QS, (i.e.
exponential service distributions and hence no fat tail), [Flatto
1997] shows that the random order service scheduling rule yields:

\begin{equation}\label{ROS-MM1}
{\rm Prob}(W_{{\rm ROS}}>x) \varpropto C_{\rho} x^{-{5 \over 6}}\,
e^{- \gamma x- \delta x^{{1\over 3}}},\,\,\,\,\,\,\,\,\,\,\,{\rm
for } \,\,\, x \rightarrow \infty,
\end{equation}

\noindent with

$$
C(\rho) = 2^{2 \over 3} 3^{-{1 \over2}} \pi^{{5\over 6}} \rho^{{17
\over 12}} {1+ \rho^{{1 \over 2}} \over  \left[1-\rho^{{1 \over
2}}\right]^{3}} \exp\left\{{1+ \rho^{{1\over2}}\over
1-\rho^{{1\over2}}}\right\},
$$

\noindent

$$
\gamma= \left(\rho^{-{1
\over2}}-1\right)^{2}\,\,\,\,\,\,\,\,\,\,\,{\rm and }
\,\,\,\,\,\,\,\,\,\, \delta =3\left[{\pi \over 2}\right]^{{2\over
3}} \rho^{-{1 \over 6}},
$$

\noindent  which has to be compared with the FCFS scheduling
discipline, which for the same $M/M/1$ QS reads as, [Cohen 1973]:

\begin{equation}\label{ROS-MM2}
{\rm Prob}(W_{{\rm FCFS}}> x)= {1 \over \beta} (1-\rho) e^{-
{1\over \beta} (1-\rho)x}.
\end{equation}

\noindent While the  detailed behaviors given by
Eq.(\ref{ROS-MM1}) and (\ref{ROS-MM2})clearly differ, they however
both share, in accord with [Barab{\'a}si 2005], an exponential decay.

\section*{Appendix  B - Deadline drawn from Pareto distribution}

\noindent Here, we focus on:

    \begin{equation}\label{PARETO}
    G(x) = \left\{
\begin{array}{l@{\qquad}l} 0 & {\rm when}\ {x \over B} < 1,\\ \,
\\ 1- \left({B \over x}\right)^{(\omega-1)}
&{\rm when}\ {x \over B} \geq 1, \qquad \omega > 1, \\
\end{array}
    \right.
\end{equation}

\noindent which has  no moment of order $\omega-1$ or higher. For
$\omega >2$,  we have $\left\langle{\cal D}\right\rangle =\left[{
\omega-1 \over \omega-2}\right]\, B$:

\begin{equation}\label{FRONTIER-1}
    {\cal F}(Q) =\left\{
\begin{array}{l@{\qquad}l}
B \left({B \lambda \over Q(\omega - 2) }\right)^{\left({1 \over
\omega-2}\right)} & {\rm when}\ {Q\over \lambda} \leq  {B \over
\omega -2},\\ \,
\\ \left({\omega-1 \over \omega-2}\right)B- {Q \over \lambda}
&{\rm when}\ {Q\over \lambda} > {B \over \omega -2}. \\
\end{array}
    \right.
\end{equation}

\noindent Using Eqs.(\ref{LTP1}) and (\ref{LTP2}), the LTP
 distribution reads as:

\begin{equation}\label{PAR1}
 {Q \over \lambda}  \geq {B \over \omega -2} \quad  \Rightarrow \quad F(x) = \left\{
\begin{array}{l@{\qquad}l}
 0, & {\rm when}\ x \leq {\cal F} (Q),\\ \,
\\ 1 - {\lambda \over Q}\left({\omega -1 \over \omega-2}B
-x\right),
&{\rm when}\ {\cal F} (Q) \leq x < B,  \\ \, \\
1- {B \lambda \over Q(\omega-2)} \left({B\over
x}\right)^{\omega-2}.& {\rm when}\ x \geq B.
\end{array}
    \right.
\end{equation}

\begin{equation}\label{PAR2}
{Q \over \lambda}  < {B \over \omega -2} \quad  \Rightarrow \quad
F(x) =\left\{
\begin{array}{l@{\qquad}l} 0, & {\rm when}\ x \leq {\cal F} (Q), \\ \,
\\ 1- {B \lambda \over Q(\omega-2)}\left(B\over x\right)^{\omega
-2}.
&{\rm when}\ x > {\cal F} (Q). \\
\end{array}
    \right.
\end{equation}

\noindent Eqs.(\ref{PAR1}) and (\ref{PAR2}) exhibit a  fat tail
with power $\omega-2$. Note that,  Eq.(\ref{PAR2}), implies
 that for ${\omega
>2 }$ and for ${Q \over \lambda} < {B \over (\omega -2)}$, the
EDF scheduling policy  part of the  tasks enter into the service
before the due date expired. Finally note also,  that for $\omega
\leq 2$, no moments exists for the deadline distribution and hence
the theory [Doytchinov et al. 2001] cannot be applied directly. We
conjecture that for these regimes no scheduling rule will be able
to deliver tasks in due time.

\vspace{1cm} \noindent

\section*{References}

\vspace{0.4cm} \noindent [Baldwin et al. 2000] R.O. Baldwin, N. J.
Davis, J. E. Kobza and S. F. Mikdiff. {\it Real-time queueing
theory : A tutorial presentation with an admission control
application}. Queueing Syst. Th. and Appl.{\bf 35}(1-4), (2000),
1-21.

 \vspace{0.4cm} \noindent [Barab{\'a}si 2005]
A.-L. Barab{\'a}si. {\it The origin of bursts and heavy tails in human
dynamics}. Nature {\bf 435}, (2005), 207.

\vspace{0.4cm} \noindent [Blanchard et al. 2004] Ph. Blanchard and
M.-O. Hongler. {\it Self-organization of critical behavior in
controlled general queueing systems}. Phys. Lett. {\bf A
323}(1-2), (2004), 63-66.

 \vspace{0.4cm} \noindent[Boxma et al. 2004]
O.J. Boxma, S.G. Foss, J.M. Lasgouttes and R. N{\`u}{\~n}ez Queija. {\it
Waiting time asymptotic in the single server queue with service in
random order},  Queueing Systems {\bf 46}, (2004), 35-73.

\vspace{0.4cm} \noindent [Brauer et al. 2001]. F. Brauer and C.
Castillo-Ch{\'a}vez. \underline{{\it Mathematical Models in}}
\underline{{\it  Population Biology and Epidemiology}}. Text in
Applied Mathematics {\bf 40}, Springer (2001).

\vspace{0.4cm} \noindent [Cohen 1973] J. Cohen. {\it Some results
on regular variation for distributions in queueing and fluctuation
theory}. J. Appl. Probab. {\bf 10}, (1973), 343-353.

\vspace{0.4cm} \noindent [Doytchinov et al. 2001] B. Doytchinov,
J. Lehozcky, S. Shreve. {\it Real-time queues in heavy traffic
with earliest-deadline-first queue discipline}. Ann. Appl. Probab.
{\bf 11}, (2001), 332-378.

\vspace{0.4cm} \noindent [Dusonchet et al. 2003] F. Dusonchet and
M.-O. Hongler. {\it Continuous-time restless bandit and dynamic
scheduling for make-to-stock production}. IEEE Trans Robot and
Autom. {\bf 19}(6), (2003) 997-990.

 \vspace{0.4cm} \noindent[Filliger et al. 2005] R. Filliger and
 M.-O. Hongler. {\it Syphon dynamics - A soluble model of multi-agents cooperative behavior}.
 Europhys. Lett.  {\bf 70}(3), (2005), 285-291.

\vspace{0.4cm} \noindent [Flatto 1977]  L. Flatto. {\it The
waiting time distribution for the random order service $M/M/1$
queue}. The Annals of Probab. {\bf 7}, (2), (1997) 382-409.

\vspace{0.4cm} \noindent [Lehozcky 1996] J. Lehozcky {\it
Real-time queueing theory}. In Proceed. of the IEEE Real-time
system symposium, New-York (1996), 186-195.

 \vspace{0.4cm} \noindent
[Lehozcky 1997] J. Lehozcky {\it Using real time queueing theory
to control lateness in realtime systems}. Perfom. Eval. {\bf 25}
(1997), 158-168.

 \vspace{0.4cm} \noindent [Pakes 1975]
A. G. Pakes. {\it On the tails of waiting-time distributions}. J.
Appl. Prob   .  {\bf 12}, (1975), 555-564.

\vspace{0.4cm} \noindent [Panwar et al. 88] S. Panwar and D.
Townsley and J.K Wolf {\it Optimal Scheduling Policies for a Class
of Queues with Customers Deadlines for the Beginning of services}.
J. of the ACM {\bf 35},(4), (1988), 832-844.

\vspace{0.4cm} \noindent [V{\'a}zquez 2005]  A. V{\'a}zquez. {\it Exact
Results for the Barab{\'a}si Model of Human Dynamics}. Physical Review
Letters {\bf 95}, (2005), 248701.

\vspace{0.4cm} \noindent [V{\'a}zquez et al 2006]  A. V{\'a}zquez, J. G.
Oliveira, Z.  Dezs{\"o}, K.-I. Goh, Kondor I. and A.-L. Barab{\'a}si. {\it
Modeling Bursts and Heavy Tails in Human Dynamics}, Phys. Rev. E
{\bf  73}(3), (2006), 036127.

 \end{document}